\documentclass[aps,prl,reprint,amsmath,amssymb,nofootinbib,floatfix]{revtex4-1}
\usepackage{graphicx} 
\usepackage{physics}
\usepackage{xcolor}
\usepackage{comment}
\usepackage{mathtools}
\usepackage{makecell}
\usepackage{hyperref}

\renewcommand{\Re}{\real}
\renewcommand{\Im}{\imaginary}
\providecommand{\ii}{\text{i}}

\newcommand{\unit}[1]{\mathbf{\hat{#1}}}
\newcommand{\bisp}{\boldsymbol{\psi}}

\begin{document}

\title{Topologies of light in electric-magnetic space}
\author{Alex J. Vernon}
\email{alex.vernon@dipc.org}
\affiliation{Donostia International Physics Center (DIPC), Donostia-San Sebasti\'an 20018, Spain}

\begin{abstract}
    In nonparaxial, monochromatic light the electric and magnetic fields generally have different energy densities, different singularities and different polarisation structures.
    A topological picture of the electric field or magnetic field in isolation cannot capture the elusive topology of nonparaxial light that exists in the spatially dependent relationship between the two fields: the degree to which light breaks fundamental symmetries (parity, duality, time-reversal).
    With this work a new ellipse is introduced that resides not in real space, but in electric-magnetic (EM) space, and whose geometry depends on these broken symmetries.
    The EM ellipse has circular and linear polarisation singularities and may be organised into particle-like textures.
    These thus-far hidden topologies are present even in rudimentary structured waves, for a second-order EM-space meron is shown to be present in a focussed linearly polarised vortex beam.     
\end{abstract}

\maketitle
Well might an electromagnetic wave, with more than one momentum space component, be embedded with singularities in many of the several vector and scalar fields that describe its physics  \cite{Nye1987,vanGogh2020,Wang2022,Lim2024}.
Singularities have a global signature in the surrounding, well-defined field that cannot be removed except by annihilation with another singularity of opposite topological charge.

With singular optics comes a more palatable view of monochromatic structured waves, not as incomprehensible, globally complex fields, but as weavings of threadlike singularities.
Even in simple configurations, these singularities tend to bring about rich, multidimensional topologies in light \cite{Sugic2021,MataCervera2025,Marco2025}.
Among the many kinds of particle-like textures that have been studied in optics \cite{Tsesses2018,Lei2021,Deng2022,Wang2024,Shen2024,Ornelas2024,Putley2025} are those of paraxial polarisation \cite{Beckley2010,Gao2020,Lin2021,Maxwell2025}; in particular Stokes skyrmions.
A beam carrying a Stokes skyrmion realises every possible polarisation ellipse, constituting a complete mapping of real space to the Poincar\'e sphere with an integer skyrmion number.
It is common for these topologies to be studied in electric quantities, like electric polarisation or electric spin angular momentum, isolated from the topology of the magnetic field (or, sometimes, vice versa).
Not unreasonably is this done in the paraxial regime, where electric and magnetic polarisation is identical except for a local $90^\circ$ rotation of every ellipse---but in nonparaxial light the two fields generally have different polarisation distributions, different energy density, and different singularities.
A polarisation ellipse is something that must be perceived by a detector with a specific sensitivity---be it towards electric fields, magnetic fields or a combination of the two.
It may be argued therefore that polarisation singularities do not have an intrinsic significance to electromagnetic waves \cite{Vernon2025}, and a more fundamental approach is to consider symmetries of light, held or broken depending on the delicate relationship \textit{between} electric and magnetic fields.
Perhaps it is possible to define an intuitive, \textit{electromagnetic} ellipse, able to describe a deeper topology of light.

Here I introduce an averaged electromagnetic ellipse that resides in `electric-magnetic' (EM) space, rather than real space, whose geometry depends on broken fundamental symmetries, and which is defined under the assumption of monochromatic waves.
Though in paraxial waves the ellipse is rather trivial, in the nonparaxial regime it reveals, for the first time, intrinsic particle-like topologies of light, including an EM-space second-order meron numerically demonstrated to lie in a tightly focussed and linearly polarised vortex beam (a meron is a half-skyrmion, e.g., a 1-to-1 mapping to only one hemisphere of the unit sphere. A second order meron can be associated with a 2-to-1 mapping to one hemisphere \cite{Krol2021}).
This work is directly linked to a Bloch sphere called the electromagnetic or energy symmetry sphere (ESS), proposed in \cite{Golat2024}.
It is also serves to develop the recently proposed ``symmetry dislocations", lines or surfaces in nonparaxial light where the field preserves a particular symmetry \cite{Vernon2025}.

The perspective offered here is meaningful not least because optical singularities have, since their discovery, been married to light-matter interactions, and by manipulating them waves can be structured to elicit certain interactions of interest.
The more exotic kinds of light-matter interactions possible in monochromatic, nonparaxial light \cite{Zhou2022,Golat2024_2}, such as discriminating chiral forces from helicity gradients \cite{Cameron2014}, dual-asymmetric reactive forces and torques \cite{NietoVesperinas2022}, and the elusive magnetoelectric interaction \cite{Bliokh2014}, result from broken symmetries in light.
If one wishes to shape light waves to purposely stimulate these exotic interactions, it would be useful to appreciate light's hidden topology in EM space.

A mathematical motivation for the EM ellipse is what was originally utilised in \cite{Bliokh2014} and developed in \cite{Golat2024}: that a monochromatic electromagnetic field $\bisp$ \cite{BialynickiBirula1996,Bliokh2014,Dennis2023},
\begin{equation}\label{bispinorEM}
    \bisp_\text{EM}=\frac{1}{2}\begin{pmatrix}\sqrt{\varepsilon_0}\mathbf{E}\\\sqrt{\mu_0}\mathbf{H}\end{pmatrix},
\end{equation}
which lives in the $\mathbb{C}^2\otimes\mathbb{C}^3\otimes\mathbb{L}^2$ vector space and where $\mathbf{E}$ and $\mathbf{H}$ are complex phasors, has a structure that resembles the Jones vector for two-dimensional electric (or magnetic) polarisation states.
To proceed further, one should be briefly familiarised with the ESS, summarised in the next section, before introducing the EM ellipse and its polarisation singularity analogues.
With this new interpretation nontrivial 2D EM-space textures are demonstrated, before concluding remarks.

\section{The energy symmetry sphere, summarised}

\begin{table}[b!]
\centering
\begin{tabular}{c|c|c|c|c|c}
    Operator & Transformation & $W_0$ & $W_1$ & $W_2$ & $W_3$\\\hline
     $\hat{P}$ & \makecell{$\sqrt{\varepsilon_0}\mathbf{E}\to-\sqrt{\varepsilon_0}\mathbf{E}$\\$\sqrt{\mu_0}\mathbf{H}\to\sqrt{\mu_0}\mathbf{H}$} & $+$ & $+$ & $-$ & $-$ \\\hline
     $\hat{D}$ & \makecell{$\sqrt{\varepsilon_0}\mathbf{E}\to\sqrt{\mu_0}\mathbf{H}$\\$\sqrt{\mu_0}\mathbf{H}\to-\sqrt{\varepsilon_0}\mathbf{E}$} & $+$ & $-$ & $-$ & $+$ \\\hline
     $\hat{T}$ & \makecell{$\sqrt{\varepsilon_0}\mathbf{E}\to\sqrt{\varepsilon_0}\mathbf{E}^*$\\$\sqrt{\mu_0}\mathbf{H}\to-\sqrt{\mu_0}\mathbf{H}^*$} & $+$ & $+$ & $-$ & $+$
\end{tabular}
\caption{Changes in the sign of Stokes-like quantities $W_{0-3}$ \eqref{Ws} under parity $\hat{P}$, discrete duality $\hat{D}$, and time reversal $\hat{T}$ operators \cite{Bliokh2014,Golat2024}.}
\end{table}
Often when dealing with paraxial polarisation it is convenient to choose a particular basis with which to represent the Jones vector, $\mathbf{E}_\text{parax}=(E_i,E_j)$, for example a horizontal-vertical basis $(E_x,E_y)$, a diagonal-antidiagonal basis $(E_\text{d},E_\text{a})$, or a right-left circular basis $(E_+,E_-)$.
It is well-known that the difference in energy of the two components of these three natural bases yield the Stokes parameters $S_{1-3}$, while the sum of the components in any basis gives $S_0$.

The bispinor \eqref{bispinorEM} was first shown to share this property in \cite{Bliokh2014}, giving rise to four Stokes-like parameters in units of energy density: total $W_0$, electric-magnetic $W_1$, magnetoelectric $W_2$ and chiral $W_3$ energy density.
These four Stokes-like parameters correspond to the differences and sums of strengths of fields $\mathbf{F}_\text{a}$ and $\mathbf{F}_\text{b}$ \cite{Golat2024}, these being the entries of the bispinor in different $\mathbb{C}^2$ bases (linear combinations of $\mathbf{E}$ and $\mathbf{H}$).
One natural $\mathbb{C}^2$ basis is that of right- and left-handed helicity eigenstates,
\begin{equation}\label{bispinorRL}
    \bisp_\text{RL}=\begin{pmatrix}\mathbf{F}_\text{R}\\\mathbf{F}_\text{L}\end{pmatrix}
    =\frac{1}{2\sqrt{2}}\begin{pmatrix}\sqrt{\varepsilon_0}\mathbf{E}+\ii\sqrt{\mu_0}\mathbf{H}\\\sqrt{\varepsilon_0}\mathbf{E}-\ii\sqrt{\mu_0}\mathbf{H}\end{pmatrix},
\end{equation}
where $\mathbf{F}_\text{R/L}$ are phasor versions of Riemann-Silberstein vectors \cite{BialynickiBirula2013}, allowing for compact definitions of $W_i$ for $i=0,1,2,3$, using the corresponding Pauli matrix $\sigma_i$ \cite{Bliokh2019},
\begin{equation}
    W_i=\bisp_\text{RL}^\dagger\cdot(\sigma_i)\bisp_\text{RL},
\end{equation}
explicitly given in the EM basis as,
\begin{equation}\label{Ws}
\begin{split}
    W_0&=\frac{1}{4}(\varepsilon_0|\mathbf{E}|^2+\mu_0|\mathbf{H}|^2),\\
    W_1&=\frac{1}{4}(\varepsilon_0|\mathbf{E}|^2-\mu_0|\mathbf{H}|^2),\\
    W_2&=-\omega h_\text{reac}=-\frac{1}{2c}\Re\{\mathbf{E}^*\cdot\mathbf{H}\},\\
    W_3&=\omega h=-\frac{1}{2c}\Im\{\mathbf{E}^*\cdot\mathbf{H}\}.
\end{split}
\end{equation}
The quantities $h$ and $h_\text{reac}$ are helicity \cite{Cameron2012,NietoVesperinas2015} and what has recently been termed reactive helicity \cite{NietoVesperinas2021,Zhang2025} densities respectively.
In connection with the standard Stokes vector, a vector $\mathbf{W}=(W_1,W_2,W_3)/W_0$ was proposed in \cite{Golat2024} along with a Bloch sphere of radius 1 within which $\mathbf{W}$ is constrained, i.e., $0\leq|\mathbf{W}|\leq1$.
Under the action of fundamental discrete symmetry operators, $W_{1-3}$ were shown either to be odd or even in \cite{Bliokh2014} (see Tab.~1), motivating the Bloch sphere to be named the energy symmetry sphere.
Just as the Poincar\'e sphere completely characterises the (paraxial) electric or magnetic polarisation ellipse, the ESS characterises the EM ellipse proposed here in this work.

\begin{figure*}[t!]
    \centering
    \includegraphics{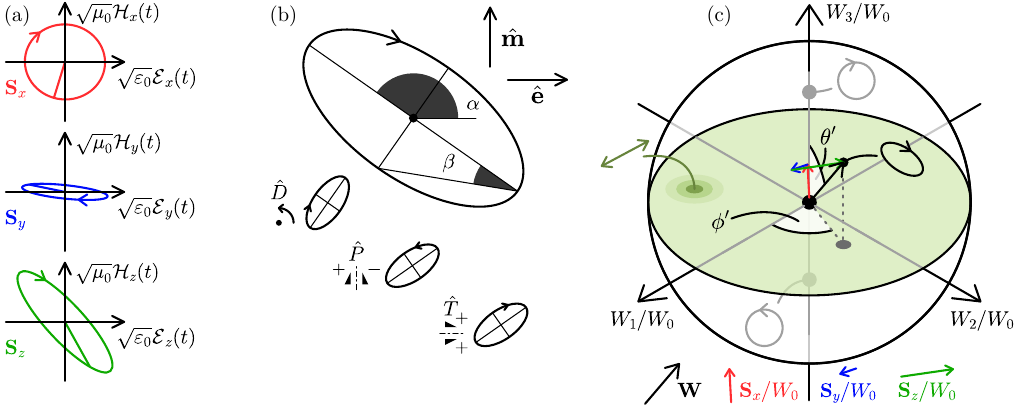}
    \caption{(a): Ellipses in EM space from the $\unit{x},\unit{y},\unit{z}$ projections of $\bisp$ at a point in a random nonparaxial field for which $\sqrt{\varepsilon_0}\mathbf{E}=(-0.11-0.39\ii,-0.36+0.24\ii,0.26-0.41\ii)$ and $\sqrt{\mu_0}\mathbf{H}=(-0.36+0.11\ii,0.08+0.03\ii,-0.46+0.23\ii)$ in arbitrary units. On each ellipse the coloured straight line is $\sqrt{\varepsilon_0}\Re{E_\text{a}}\unit{e}+\sqrt{\mu_0}\Re{H_\text{a}}\unit{m}$, indicating the phase. Each ellipse has an EM Stokes vector $\mathbf{S}_\text{a}=(S_{\text{a}1\text{(EM)}},S_{\text{a}2\text{(EM)}},S_{\text{a}3\text{(EM)}})$. (b): The overall EM ellipse made by a weighted average of the Stokes parameters of the ellipses in (a), with an orientation angle $\alpha$ (with respect to $\unit{e}$) and angular eccentricity $\beta$.
    Operators $\hat{D},\hat{P},\hat{T}$ transform the ellipse: $\hat{D}$ is a $90^\circ$ anticlockwise rotation, $\hat{P}$ is a reflection, $\hat{T}$ is a reflection that preserves handedness.
    (c): The energy symmetry sphere \cite{Golat2024} and its correspondence with the EM ellipse in (b).
    The azimuth and elevation angles are [due to the sign convention of \eqref{Ws}] $\phi'=-2\alpha$ and $\theta'=\pi/2+2\beta$.}
    \label{fig1}
\end{figure*}

\section{The EM ellipse}
It is useful to represent the monochromatic bispinor \eqref{bispinorEM} with a single-line expression,
\begin{equation}\label{bispinor_linear}
    \bisp_\text{EM}=\frac{\sqrt{\varepsilon_0}}{2}\mathbf{E}\otimes\unit{e}+\frac{\sqrt{\mu_0}}{2}\mathbf{H}\otimes\unit{m},
\end{equation}
where the unit vectors $\unit{e}$ and $\unit{m}$ that belong to the $\mathbb{C}^2$ vector space differentiate the rows of the bispinor in \eqref{bispinorEM}.
The conventional electric or magnetic polarisation ellipse is a projection of the bispinor \eqref{bispinor_linear} onto one of $\unit{e}$ or $\unit{m}$, whose time-dependent trajectory is plotted in real space---the space spanned by three real-space basis vectors $\unit{u}_\text{a}$ for $\text{a}=1,2,3$.
The polarisation ellipse is an effective visualisation tool, but is nevertheless an obscured view of the whole electromagnetic field, obtained by generally discarding half of the bispinor's components.
It is only constructed from the relationship between scalar components internal to $\mathbf{E}$, or to $\mathbf{H}$, neglecting the relation of the components of $\mathbf{E}$ to their counterparts in $\mathbf{H}$.

There is no reason why one cannot instead project $\bisp$ onto the orthonormal real-space basis vectors, $\bisp\cdot\unit{u}_\text{a}$, and plot one of three ellipses in the \textit{two}-dimensional space spanned by $\unit{e}$ and $\unit{m}$---the EM space---as is shown for a point in a nonparaxial field in Fig.~\ref{fig1}(a).
The three projections of the ellipses, in this case onto $\unit{u}_\text{a}=\unit{x},\unit{y},\unit{z}$, each have an orientation, ellipticity and well-defined phase in EM space.
Each ellipse has a Stokes vector $\mathbf{S}_\text{a}$ determined by four EM Stokes parameters:
\begin{equation}\label{EM_stokes}
    \begin{split}
        S_\text{a0(EM)}&=(\varepsilon_0|E_\text{a}|^2+\mu_0|H_\text{a}|^2)/4,\\
        S_\text{a1(EM)}&=(\varepsilon_0|E_\text{a}|^2-\mu_0|H_\text{a}|^2)/4,\\
        S_\text{a2(EM)}&=-\Re\{E_\text{a}^*H_\text{a}\}/(2c),\\
        S_\text{a3(EM)}&=-\Im\{E_\text{a}^*H_\text{a}\}/(2c),
    \end{split}
\end{equation}
satisfying $S_\text{a0(EM)}^2=S_\text{a1(EM)}^2+S_\text{a2(EM)}^2+S_\text{a3(EM)}^2$ [with sign convention and prefactors matching \eqref{Ws}] and which are in general different for each $[\text{a}]$.
Summing the $0^\text{th}$ parameter in \eqref{EM_stokes} over [a] gives the total energy density $W_0$, and likewise $\sum_\text{a} S_{\text{a}i\text{(EM)}}=W_i$.
Therefore the vector $\mathbf{W}=(W_1,W_2,W_3)/W_0$ can be interpreted as an energy-density-weighted average over [a] of the parameters $1-3$ in \eqref{EM_stokes}.
The vector $\mathbf{W}$ can be used to construct an averaged ellipse that maps to the ESS---the EM ellipse, the topic of this work---which is shown in Fig.~\ref{fig1}(b).
Figure \ref{fig1}(a)'s three ellipses are visibly different to each other for the $\mathbf{E}$ and $\mathbf{H}$ phasors given in the caption such that the EM ellipse maps to a point inside the ESS in Fig.~\ref{fig1}(c).
Were the electric and magnetic fields related by a complex constant $\gamma$, $\sqrt{\varepsilon_0}\mathbf{E}=\gamma\sqrt{\mu_0}\mathbf{H}$, the three ellipses would have the same shape and orientation in Fig.~\ref{fig1}(a) and the EM ellipse would map to a point on the surface of the ESS.
Examples of this situation could include an electric field dark spot $\mathbf{E=0}$, for which each projection is vertically polarised along $\unit{m}$, or a right-handed pure helicity state for which $\mathbf{E}=\ii\eta\mathbf{H}$, and each projection is clockwise-circularly polarised in EM space.

The EM ellipse's handedness is given by the sign of $W_3$, its angular eccentricity by $\beta=-(1/2)\arctan{W_3/\sqrt{W_1^2+W_2^2}}$, and its orientation (subtended by its semi-major axis and $\unit{e}$) in EM space by $\alpha=-(1/2)\arg{\mathbf{F}_\text{R}^*\cdot\mathbf{F}_\text{L}}$.
According to Tab.~1, the symmetry operators $\hat{D},\hat{P},\hat{T}$ can be interpreted as rigid transformations of the EM ellipse: discrete duality $\hat{D}$ rotates the ellipse through $90^\circ$ anticlockwise; parity $\hat{P}$ reflects the ellipse (via $\unit{e}\to-\unit{e}$); time-reversal $\hat{T}$ also reflects the ellipse (via $\unit{m}\to-\unit{m}$) but without changing its handedness (there is, in fact, an analogous symmetry behaviour among the traditional Stokes parameters \cite{Bliokh2015}).
Invariance to any one of these transformations indicates a symmetry of the field.
Alternatively stated, when the EM ellipse is defined for a point in space (i.e., as long as $\mathbf{W}\neq\mathbf{0}$), the field must be asymmetric to some degree because the EM ellipse cannot be invariant to all individual operations in Tab.~1 simultaneously.
The meaning of the EM ellipse is, therefore, that its geometry represents broken fundamental symmetries of light.

It must be strongly emphasised that the EM ellipse exists in the 2D space spanned by $\unit{e}$ and $\unit{m}$, not in real-space.
It remains two-dimensional regardless of paraxiality, mapping to the ESS as in Fig.~\ref{fig1}(c) and always consistently parametrised by $W_i$ in monochromatic light even when in the nonparaxial regime the traditional electric and magnetic polarisation ellipses, in contrast, become more challenging to characterise due to their general 3D orientation.
Although in paraxial waves, where $W_1=W_2=0$, the EM ellipse is always either circular or undefined depending on the value of helicity density, it may be organised into topologically nontrivial distributions in nonparaxial waves.

The intrinsic meaning of the EM ellipse set out above, and the meaning of a traditional electric or magnetic polarisation ellipse, are worth comparing.
For the EM ellipse is \textit{completely basis-independent} while polarisation ellipses are not \cite{Vernon2025}.
Basis-dependence can be understood by initially imagining an illuminated nanorod in vacuum, which couples to field components along its axis. 
If the nanorod is aligned to the $x$ axis then, breaking the symmetry of the system, it is most sensitive to the $\unit{x}$ component of the electric field.
Of course, $E_x$ has no general importance in light because the nanorod could simply be rotated and couple to $E_y$, or be replaced by a particle with a different symmetry.
It is not only the geometry of the nanorod that breaks the system's symmetry---its material does, too, for it has been implicitly assumed (by saying it couples to $E_x$) that the rod is electrically polarisable, $\varepsilon\neq1$, while having vacuum permeability, $\mu\approx1$.
The rod is strongly sensitive to the $\unit{e}$ component of the bispinor in \eqref{bispinor_linear}, yet, much like rotating the rod in real space to change the cartesian field component being detected, the rod's coupling in EM space can be rotated by changing its material.
A magnetically polarisable material couples to $\mathbf{H}$, while a material with both $\varepsilon\neq1$ and $\mu\neq1$ couples to some linear combination of $\mathbf{E}$ and $\mathbf{H}$.
Polarisation, therefore, is a $\mathbb{C}^2$ basis-dependent property that is perceived differently depending on the sensitivity of a detector.
But the meaning of an EM ellipse of a particular shape does not rely on the presence of a particular detector to be justified, for it corresponds to the behaviour of the electromagnetic field under $\hat{D},\hat{P},\hat{T}$ operators.
Whereas the apparent position of polarisation singularities (C lines, L lines \cite{Nye1987}) in nonparaxial light depends on the detector (e.g., an electric field detector, or a magnetic field detector), the position of singularities of the EM ellipse does not.

\section{Singularities of the EM ellipse}
\begin{figure}[t!]
    \centering
    \includegraphics[width=\columnwidth]{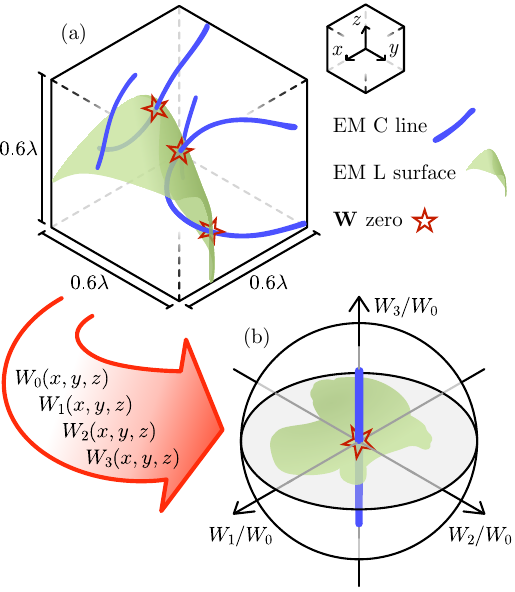}
    \caption{C lines (blue) and L surfaces (green) of the EM ellipse in a random nonparaxial light field formed of seven plane waves, plotted in (a) real space and in (b) the ESS ($\mathbf{W}$ space).
    The C lines map in (b) to a portion of the $W_3$ axis in the ESS, while the L surface covers a portion of the $W_3=0$ plane within the ESS.
    Where a C line punctures an L surface  the vector $\mathbf{W}$ is zero.
    There are three zeros of $\mathbf{W}$ in (a), shown by the red-outlined stars, and all three map to the centre of the ESS in (b).}
    \label{fig2}
\end{figure}

Symmetry-based singularities were proposed in ref. \cite{Vernon2025} where light is invariant to one or a combination of $\hat{D},\hat{P},\hat{T}$ operators.
These `symmetry dislocations' can also be understood as extremes of the EM ellipse: the ellipse is invariant to $\hat{D}$ when it is circular, and invariant to $\hat{P}\hat{T}$ when it is linear.
A circular EM ellipse corresponds to an ill-defined orientation $\alpha$, meaning $\mathbf{F}_\text{R}^*\cdot\mathbf{F}_\text{L}=W_1+\ii W_2=0$ (codimension 2, satisfied along lines also recognised as time-averaged Riemann-Silberstein vortices \cite{BialynickiBirula2003,Kaiser2004,Berry2004}), while a linear EM ellipse has $W_3=0$ (codimension 1, satisfied on surfaces).
EM-space C lines and L surfaces are shown in Fig.~\ref{fig2} for a nonparaxial superposition of seven random plane waves, in (a) as they appear in real space, and in (b) as they respectively map to the $W_3/W_0$ axis and $W_3/W_0=0$ plane in the ESS (details of the seven plane waves producing the field in Fig.~\ref{fig2} specifically are found in the supplementary information).

When an EM-space C line pierces an EM-space L surface (red stars in Fig.~\ref{fig2}), the vector $\mathbf{W}$ vanishes and the ellipse is entirely undefined.
The natural existence of this kind of point-defect contrasts with traditional polarisation singularities, C and L lines of the electric field, which do not generically intersect (unless forced to when the field is zero \cite{Vernon2023}).

The shape and orientation of the EM ellipse that corresponds to a given azimuth and elevation angle of $\mathbf{W}$ within the ESS is analogous to the relationship between 2D polarisation and the Poincar\'e sphere.
It is natural to ask whether it is possible to achieve a complete mapping of all possible ellipses from real space to the ESS and realise a topologically nontrivial distribution, akin to a skyrmion.

\section{Particle-like textures}

This section proposes an intrinsic particle-like texture of light that has thus far been hidden in EM space.
It arises from the mapping of an electromagnetic field to the ESS and are in this sense analogous to Stokes skyrmions/merons \cite{Beckley2010,Gao2020,Marco2024}, except that nontrivial EM-space textures exist only in nonparaxial light where $W_1$ and $W_2$ can be non-zero.
These EM space textures rather satisfyingly combine the intuitive, Poincar\'e-like classification of Stokes topologies with the nonparaxial generality of near-field, instantaneous field \cite{Tsesses2018} and spin skyrmions \cite{Lei2021}.

Such a particle-like texture can be identified by the winding number of the vector $\mathbf{W}$ over a plane $S$ in real space, which measures the wrapping of real space over the azimuth and elevation angles of the ESS.
This winding number, or skyrmion number, can be calculated from \cite{Shen2024}
\begin{equation}\label{skyrmion_number}
    n=\frac{1}{4\pi}\int_S\mathbf{w}\cdot\left(\frac{\partial\mathbf{w}}{\partial u}\times\frac{\partial\mathbf{w}}{\partial v}\right)dudv,
\end{equation}
where $\mathbf{w}=\mathbf{W}/|\mathbf{W}|$.
Equation \ref{skyrmion_number} also functions as a general winding number for a vector over an arbitrary surface $S$, parametrised by the two general variables $u,v$.
It is straightforward to argue that, although they are not flattened over a two-dimensional plane, skyrmion-like features are anchored to the naturally occurring point-zeros of $\mathbf{W}$ in nonparaxial waves.
Over any zero-enclosing surface $\mathbf{W}$ is forced to wind an integer number of times, producing every possible EM ellipse.

\begin{figure}
    \centering
    \includegraphics[width=\columnwidth]{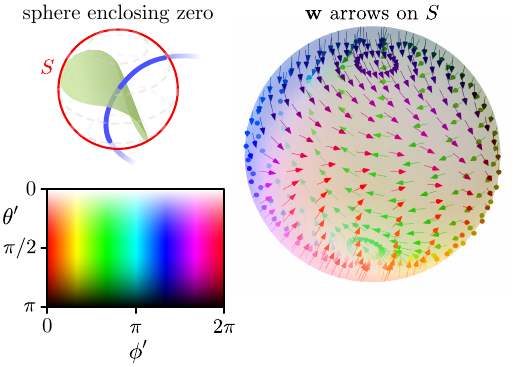}
    \caption{The vector texture that accompanies a zero of the vector $\mathbf{W}$.
    A sphere $S$ is constructed around the central zero of Fig.~\ref{fig2} [the middle red outlined star of Fig~\ref{fig2}(a)], and the normalised vector $\mathbf{w}=\mathbf{W}/|\mathbf{W}|$ is plotted on its surface (note that $\mathbf{w}$ points someway in $\mathbf{W}$ space, but its direction is visualised in real space by interpreting the $w_1,w_2,w_3$ components as $x,y$ and $z$ components).
    Arrows are assigned a colour depending on the azimuth and elevation angles ($\phi',\theta'$) of the corresponding point in the ESS.
    The winding number, calculated over $S$, is $n=-1$.}
    \label{fig3}
\end{figure}

\begin{figure*}[t!]
    \centering
    \includegraphics{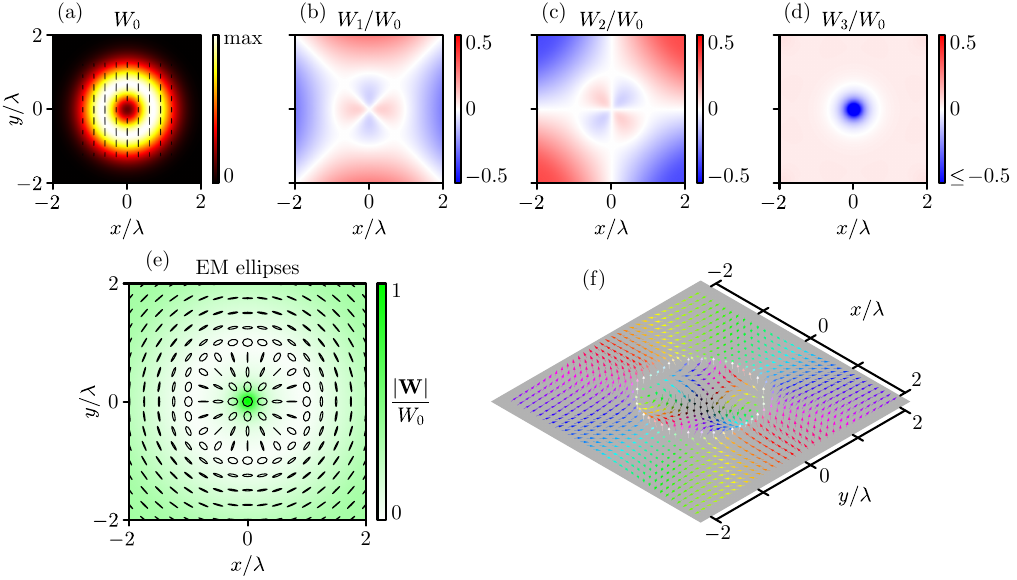}
    \caption{A particle-like EM-space texture in the focal plane $z=0$ of a focussed, $\unit{y}$-polarised vortex beam.
    The topological charge of the beam is $l=1$ and the beam waist is $1\lambda$.
    (a): the beam energy density $W_0$, with black transverse electric polarisation ellipses drawn on top, and (b)-(d) the other three Stokes-like parameters $W_{1-3}$, normalised by energy density.
    In (d), the blue end of the colour bar is saturated at $-0.5$ (exactly at the beam centre, $W_3/W_0=-1)$.
    (e): EM ellipses in the focal plane, constructed from $W_{1-3}$, while the background colour measures the degree of asymmetry of the field via the length $|\mathbf{W}|$.
    (f): normalised vector $\mathbf{w}=\mathbf{W}/|\mathbf{W}|$ plotted across the focal plane (the colour assignment for each arrow is the same as in Fig.~\ref{fig3}).
    Integrating over the whole focal plane (beyond the $x$ and $y$ limits of the plots) the winding number of this vector \eqref{skyrmion_number} is $1$.}
    \label{fig4}
\end{figure*}

Taking the middle zero of $\mathbf{W}$ of the three in Fig.~\ref{fig2}(a), where an EM-space C line intersects with the L surface, and enclosing it in a spherical surface $S$ (as shown in Fig.~\ref{fig3}), the winding number can be determined by \eqref{skyrmion_number} with $u,v=\theta,\phi$ being the real-space azimuth and elevation angles on $S$.
In Fig.~\ref{fig3} the normalised vector $\mathbf{w}$ is plotted on the enclosing sphere and is shown to realise all 3D orientations, producing a winding number of $n=-1$.
The colour of each arrow is assigned depending on the azimuth and elevation angles $\theta',\phi'$ of $\mathbf{w}$ in the ESS [in $\mathbf{W}$ space, see Fig.~\ref{fig1}(c)].
Unless another zero of $\mathbf{W}$ is enclosed by $S$ a continuous change to its size or shape does not affect the value of the topological invariant $n$.
Because of the stablility of the zero $\mathbf{W=0}$ (that it occurs naturally in nonparaxial light), the vector texture it imprints is also robust to perturbations as can be seen in the supplementary information, where the addition of a plane wave to the field simply displaces---rather than destroys---the $\mathbf{W}$ zero.
Despite the integer winding number $n$, this texture is, of course, not a true skyrmion because $n$ is associated with the point defect $\mathbf{W=0}$.

An example of a particle-like EM-space texture in absence of a point defect, a second order meron, can be demonstrated in the cross section of a linearly polarised, strongly focussed vortex beam.
That the beam is linearly polarised---in this case, the electric field polarised along $\unit{y}$, and the beam has a topological charge $l=1$---and therefore without cylindrical symmetry (unlike radial, azimuthal and circularly polarised vortex beams), appears to be important for generating a non-zero value of $W_2$ (as was recently studied \cite{Zhang2025}).
There are different approaches to numerically modelling a focussed vortex beam; an open-source angular spectrum integration method \cite{KingsleySmith2023} is used here, giving a focussed beam that is an exact solution to Maxwell's equations, but the reported results have been verified in parallel using a first-order approximation of longitudinal components (the approach of, e.g., \cite{Green2023}, which is used to generate results in the SI, showing that the texture exists largely because of longitudinal field components).

The quantities $W_i$ \eqref{Ws} were calculated in the focal plane ($z=0$) of the beam and are plotted in Fig.~\ref{fig4}(a-d), where $W_1$, $W_2$, and $W_3$ are normalised by $W_0$ and take on a value between $\pm1$ (in Fig.~\ref{fig4} the colorbar of (b)-(d) is capped at $\pm0.5$ for better contrast).
While the beam's energy density (a) retains the familar, rotationally symmetric doughnut distribution, the quantities $W_1$ and $W_2$ have four-lobed distributions offset by a 45 degree rotation.
Both $W_1$ and $W_2$ are zero at the centre of the beam and in a $2\lambda$-diameter ring.
Meanwhile the third Stokes-like parameter, $W_3$, is negative in the centre and crosses zero in the radial direction.
This is consistent with previous studies of chirality density (proportional to $W_3$ in monochromatic light) in linearly polarised focussed vortex beams and its relation to the beam's topological charge (see \cite{Forbes2021} Fig.~4).

The offset azimuthal distribution of $W_1$ and $W_2$ resembles that of the Stokes parameters $S_1$ and $S_2$ in a second-order meron \cite{Krol2021} with winding number $|n|=1$ (first-order merons have $|n|=1/2$).
However, in Fig.~\ref{fig4}(c), $W_3$ changes sign at a certain radial distance within the white zero rings of $W_1$ and $W_2$ present in Fig.~\ref{fig4}(b) and (c).
This means that when integrating within the $\lambda$ radius of the white zero rings in Fig.~\ref{fig4}(b) and (c), the winding number \eqref{skyrmion_number} (with $u,v=x,y$) approaches a value of $2$---however, beyond this radius, the winding sense of $\mathbf{w}$ reverses and, because in this region the sign of $W_3$ is strictly positive, unwraps the texture from the upper hemisphere of the ESS.
Then, integrating \eqref{skyrmion_number} over the whole focal plane produces a winding number of $1$, due to the double-wrapping of angles of $\mathbf{w}$ within the lower hemisphere of the ESS near the centre of the focal plane (where in Fig.~\ref{fig4}(c) $W_3$ is negative).
The particle-like, EM-space texture of the linearly polarised vortex beam can thus be identified as a second-order meron.

Figure \ref{fig4}(e) shows how the EM ellipse varies across the focal plane (the ellipses are drawn so that $\unit{e}$ and $\unit{m}$ correspond to the $\unit{x}$ and $\unit{y}$ directions).
In Fig.~\ref{fig4}(f), the normalised vector $\mathbf{w}=\mathbf{W}/|\mathbf{W}|$ is plotted over the focal plane (where $W_1,W_2$ and $W_3$ respectively correspond to the $x,y$ and $z$ components of the vectors as they are drawn in the figure).

While the EM ellipse topology of light has been framed in this work as an `intrinsic' property, compared to its polarisation topology that depends on the basis in $\mathbb{C}^2$ of \eqref{bispinorEM}, it may be challenging to measure.
In particular, the direct detection of $W_2$ requires a non-reciprocal particle, sensitive to magnetoelectric effects \cite{Bliokh2014}.
Equally, the quantities in Fig.~\ref{fig4}(b)-(d) have been normalised by the total energy density $W_0$---which is very low beyond the bright doughnut of Fig.~\ref{fig4}(a), meaning only a portion of the full $n=1$ meron texture of the beam could be realistically measured in this particular configuration.
It should be emphasised that this is the first proposal of and EM ellipse texture of any kind---it may be possible to realise particle-like EM ellipse textures in other configurations, such as evanescent wave superpositions \cite{Tsesses2018,Lei2021}.

\section{Discussion}

In nonparaxial, monochromatic light the electromagnetic field tends to break fundamental symmetries: the energy carried by the $\mathbf{E}$ and $\mathbf{H}$ fields is generally spread differently over space (broken duality symmetry); the combined electric and magnetic field geometry can be locally chiral (broken parity symmetry); the field can break time-reversal symmetry.
Meanwhile the electric and magnetic polarisation structures are distinct, containing different singularities, which raises the question of how one can appreciate the topological properties of light as a whole, rather than the isolated topologies of $\mathbf{E}$ and $\mathbf{H}$.

Broken symmetries of light can be interpreted using a Bloch sphere called the energy symmetry sphere \cite{Golat2024}, which can be thought of as a Poincar\'e sphere for the electromagnetic bispinor \eqref{bispinorEM}.
That the bispinor shares a similar structure to the Jones vector for paraxial polarisation motivated the introduction in this work of an EM ellipse that resides not in real space, but in electric-magnetic space.
With this ellipse, one reveals that light contains a hidden EM-space texture brought about by broken fundamental symmetries.
Naturally, this EM-space structure contains singularities, which in nonparaxial light include EM-space C lines and L surfaces, and point-like singularities where the EM ellipse is totally undefined.
It is possible for particle-like textures of the EM ellipse to exist; a second-order meron was numerically shown to be be hidden in a linearly polarised, focussed vortex beam.

It is hoped that this new perspective on light's topology in terms of the relationship between electric and magnetic fields would be useful for structuring waves to elicit special light-matter interactions characterised by the energy densities $W_{1,2,3}$.
At the same time, optical skyrmions and merons have been proposed as a means for robust information storage \cite{Yang2025}, and to this end the possibility of creating particle-like EM-space textures could provide more degrees of freedom.
While these EM-space textures are a strictly nonparaxial, nonpropagating phenomenon, it may be possible to confine them to a 2D interface via evanescent wave superpositions \cite{Tsesses2018,Lei2021}.

\section{Acknowledgments}
I thank Dr. S. Golat and Dr. F. J. Rodr\'iguez-Fortu\~no for the conversations that inspired this work.
I also thank Dr. F. J. Rodr\'iguez-Fortu\~no and Prof. Konstantin Y. Bliokh for helpful feedback on the manuscript.

\bibliography{bibliography}

\end{document}


\title{Supplementary Information for `Topologies of light in electric-magnetic space'}

\author{Alex J. Vernon}
\email{alex.vernon@dipc.org}
\affiliation{Donostia International Physics Center (DIPC), Donostia-San Sebasti\'an 20018, Spain}

\maketitle
This document supplements the main article by exploring the stability of the singularties in Fig.~2 in the manuscript, and providing results of the second-order meron texture in the linear polarised vortex beam (Fig.~4) with an alternative mathematical model.

\section{Perturbing the singularties in Fig.~2 of the main text}
\begin{figure*}[t]
    \centering
    \includegraphics{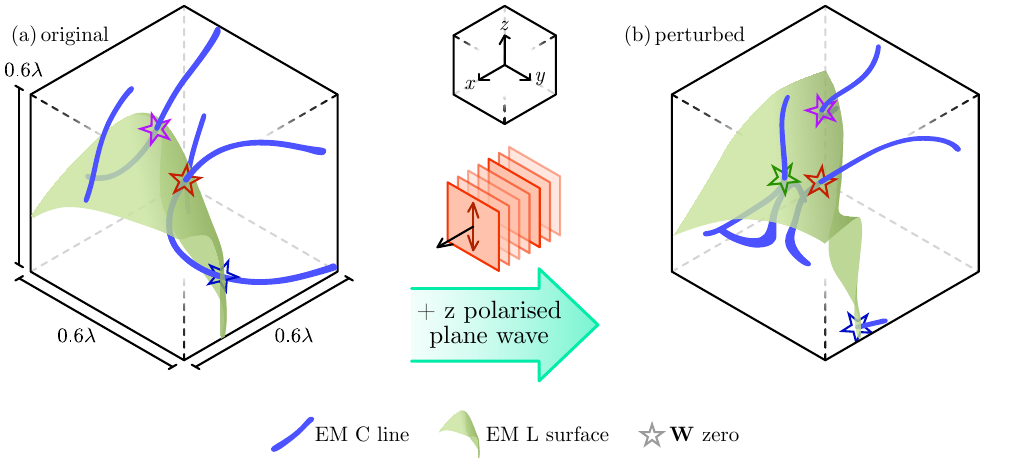}
    \caption{Behaviour of zeros of $\mathbf{W}$ after disturbance by a linearly polarised plane wave. (a) shows the original field (identical to Fig.~2(a) of the main article) now with coloured stars differentiating each zero of $\mathbf{W}$; (b) shows the perturbed field after the addition of a $\unit{z}$-polarised plane wave propagating along $x$.
    The zeros are shown to be displaced in (b) from (a), not destroyed, and a fourth zero (the green star) enters the simulation domain.}
    \label{figS1}
\end{figure*}

Figure 2 of the main text was generated by interfering seven random monochromatic plane waves.
The total field developed by the seven-plane-wave superposition was,
\begin{equation}
\begin{split}
    \mathbf{E}_\text{tot}(\mathbf{r})&=\sum_{n=1}^7\mathbf{E}_\text{n}\ee^{\ii\mathbf{k}_\text{n}\cdot\mathbf{r}},\\
    \mathbf{H}_\text{tot}(\mathbf{r})&=\sqrt{\frac{\varepsilon_0}{\mu_0}}\sum_{n=1}^7\unit{k}_\text{n}\times\mathbf{E}_\text{n}\ee^{\ii\mathbf{k}_\text{n}\cdot\mathbf{r}},
\end{split}
\end{equation}
where, in arbitrary units,
\begin{align}
    \mathbf{E}_1&=(-0.7248+0.3678\ii)\unit{x}+(0.1054 - 0.9460\ii)\unit{y}+(-0.4750-0.0839\ii)\unit{z},\\
    \mathbf{E}_2&=(0.4197 - 0.4303\ii)\unit{x}+(-0.5643 + 0.6162\ii)\unit{y}+(-0.7126 + 0.7888\ii)\unit{z},\\
    \mathbf{E}_3&=(0.3320 + 0.3763\ii)\unit{x}+(-0.5043 + 0.4339\ii)\unit{y}+(0.5751 - 0.1258\ii)\unit{z},\\
    \mathbf{E}_4&=(0.5678 - 0.1971\ii)\unit{x}+(-0.6560 + 0.9284\ii)\unit{y}+(0.1954 - 0.1113\ii)\unit{z},\\
    \mathbf{E}_5&=(-0.0255 - 0.1716\ii)\unit{x}+(1.0186 - 0.3052\ii)\unit{y}+(-0.1989 + 0.2496\ii)\unit{z},\\
    \mathbf{E}_6&=(-0.3717 + 0.2213\ii)\unit{x}+(0.4954 - 0.1265\ii)\unit{y}+(0.1201 - 0.4500\ii)\unit{z},\\
    \mathbf{E}_7&=(0.9584 - 0.1048\ii)\unit{x}+(0.1609 - 0.0430\ii)\unit{y}+(-0.2689 - 0.2238\ii)\unit{z},
\end{align}
and, with wavenumber $k=2\pi/\lambda$,
\begin{align}
    \mathbf{k}_1&=( -0.5540\unit{x}-0.2847\unit{y}+0.7823\unit{z})k,\\
    \mathbf{k}_2&=(-0.2032\unit{x}-0.8223\unit{y}+0.5315\unit{z})k,\\
    \mathbf{k}_3&=( 0.4034\unit{x}-0.5598\unit{y}-0.7238\unit{z})k,\\
    \mathbf{k}_4&=(-0.2625\unit{x}+0.0597\unit{y}+0.9631\unit{z})k,\\
    \mathbf{k}_5&=(-0.7191\unit{x}-0.1505\unit{y}-0.6784\unit{z})k,\\
    \mathbf{k}_6&=( 0.8033\unit{x}+0.5441\unit{y}+0.2420\unit{z})k,\\
    \mathbf{k}_7&=(0.1914\unit{x}-0.9766\unit{y}+0.0979\unit{z})k.
\end{align}
The conditions $\mathbf{F}_\text{R}^*\cdot\mathbf{F}_\text{L}=0$ [see main text, Eq.~(3)] and $\Im\{\mathbf{E}^*\cdot\mathbf{H}\}=0$ locate C lines and L surfaces of the EM ellipse respectively, as seen in Fig.~\ref{figS1}(a), where the zeros of $\mathbf{W}$ are differentiated by coloured stars.
Adding an eighth plane wave to to the superposition with polarisation vector $\mathbf{E}_8=\unit{z}$ and wavevector $\mathbf{k}_8=k\unit{x}$ perturbs the singularities in Fig.~\ref{figS1}(a), including the zeros of $\mathbf{W}$, but does not destroy them as seen in Fig.~\ref{figS1}(b)---rather, they are displaced.
A fourth zero (the green star) also enters the simulation domain due to the perturbation.

\section{Alternative model of the linearly polarised focussed vortex beam}

The $\unit{y}$ polarised, focussed $l=1$ vortex beam in Fig.~4 of the main article was generated using an open-source code, BEAMS \cite{KingsleySmith2023}, which produces an exact solution to Maxwell's equations for a range of focussed beams.
One alternative approach numerically simulating a focussed vortex beam is presented in \cite{Forbes2021}, which begins with the closed-form expressions of paraxial electric and magnetic fields of an optical vortex beam, initially neglecting longitudinal field components.
These expressions are solutions to Maxwell's equations in the paraxial approximation only, but an exact nonparaxial solution can be approached by iteratively calculating longitudinal field components using Gauss' laws, followed by `higher-order' transverse field components (which arise due to focussing) via Faraday's law and the Maxwell-Ampere law (and repeating the process).

Focussing effects are often sufficiently modeled by truncating this process at the first iteration, i.e., obtaining a first-order nonparaxial approximation by keeping the transverse electric and magnetic fields $\mathbf{E}_\text{t}$ and $\mathbf{H}_\text{t}$ equal to their paraxial expressions, while approximating longitudinal field components from Gauss' laws for $\mathbf{E}=\mathbf{E}_\text{t}+E_z\unit{z}$ and $\mathbf{H}=\mathbf{H}_\text{t}+H_z\unit{z}$ (taking the beam propagation along $z$).
Explicitly, this means:
\begin{equation}\label{long_approx}
    E_z\approx\frac{\ii}{k}\nabla_\text{t}\cdot\mathbf{E}_\text{t},\quad H_z\approx\frac{\ii}{k}\nabla_\text{t}\cdot\mathbf{H}_\text{t},
\end{equation}
For a paraxial Laguerre-Gauss beam of topological charge $l$, order $p$, these transverse field vectors are given by \cite{Green2023},
\begin{equation}\label{transverse_components}
    \mathbf{E}_\text{t}=E_0(\alpha\unit{x}+\beta\unit{y})f_\text{LG},\quad\mathbf{H}_\text{t}=\frac{E_0}{\eta}(-\beta\unit{x}+\alpha\unit{y})f_\text{LG},
\end{equation}
where $E_0$ is an amplitude scale factor, $\eta=\sqrt{\mu_0/\varepsilon_0}$, $\alpha$ and $\beta$ are complex constants determining the polarisation of the beam such that $|\alpha|^2+|\beta|^2=1$, and $f_\text{LG}$ is given by,
\begin{equation}
    f_\text{LG}(\rho,\phi,z)=\sqrt{\frac{2p!}{\pi w_0^2[p+|l|!]}}\frac{w_0}{w(z)}\left[\frac{\sqrt{2}\rho}{w(z)}\right]^{|l|}L_p^{|l|}(a)\ee^{\frac{-\rho^2}{w^2(z)}}\ee^{\ii\left[kz+l\phi+\frac{k\rho^2}{2R(z)}-[2p+|l|+1]\zeta(z)\right]}.
\end{equation}
In this expression, $w_0$ is the beam waist in the focal plane, $w(z)=w_0\sqrt{1+z^2/z_\text{R}^2}$ for which the Rayleigh range is $z_\text{R}=kw_0^2/2$, the wavefront curvature is $R(z)=z-z_\text{R}^2/z$, and $L_p^{|l|}(a)$ is the generalised Laguerre function with argument $a=2\rho^2/w^2(z)$.
The function $\zeta(z)=\arctan z/z_\text{R}$ is the Gouy phase.

Longitudinal components, calculated with \eqref{long_approx}, are given explicitly in \cite{Green2023} as,
\begin{equation}\label{longitudinal_components}
\begin{split}
    E_z&=\frac{\ii E_0}{k}[(\alpha\gamma+\ii\beta l/\rho)\cos\phi+(-\ii\alpha l/\rho+\beta\gamma)\sin\phi]f_\text{LG},\\
    H_z&=\frac{\ii E_0}{\eta k}[(\ii\alpha l/\rho-\beta\gamma)\cos\phi+(\alpha\gamma+\ii\beta l/\rho)\sin\phi]f_\text{LG},
\end{split}
\end{equation}
where,
\begin{equation}\label{gamma}
    \gamma=\frac{|l|}{\rho}-\frac{2\rho}{w^2(z)}+\frac{ik\rho}{R(z)}-\frac{4\rho L_{p-1}^{|l|+1}(a)}{w^2(z)L_p^{|l|}(a)}.
\end{equation}
\begin{figure*}[t]
    \centering
    \includegraphics{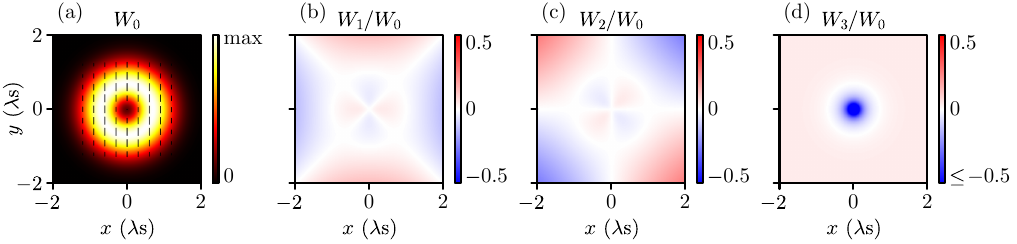}
    \caption{Distributions of $W_i$ in \eqref{SIWs} in a $\unit{y}$-polarised, $l=1$ vortex beam calculated using a first-order approximation of longitudinal field components, \eqref{transverse_components} to \eqref{gamma}.
    The electric field polarisation is shown by the black lines in (a).}
    \label{figS2}
\end{figure*}
The focal plane of the vortex beam in Fig.~4 of the main article can be approximated using \eqref{transverse_components} to \eqref{gamma} by choosing $\alpha=0$, $\beta=1$; and setting $w_0=\lambda$, $l=1$, $p=0$.
The distribution of the Stokes-like quantities,
\begin{equation}\label{SIWs}
\begin{split}
    W_0&=\frac{1}{4}(\varepsilon_0|\mathbf{E}|^2+\mu_0|\mathbf{H}|^2),\\
    W_1&=\frac{1}{4}(\varepsilon_0|\mathbf{E}|^2-\mu_0|\mathbf{H}|^2),\\
    W_2&=-\frac{1}{2c}\Re\{\mathbf{E}^*\cdot\mathbf{H}\},\\
    W_3&=-\frac{1}{2c}\Im\{\mathbf{E}^*\cdot\mathbf{H}\},
\end{split}
\end{equation}
in the beam's focal plane at $z=0$ can then be calculated, and are plotted in Fig.~\ref{figS2}(a)-(d).
Their distributions are almost the same as in Fig.~4 of the main article, except for $W_{1-3}/W_0$ (b)-(d) being paler in colour under the same colourbar (values capped between $\pm0.5$).
This is due to the lack of higher-order transverse components of $\mathbf{E}$ and $\mathbf{H}$; by recalculating $\mathbf{E}_\text{t}$ and $\mathbf{H}_\text{t}$ using Faraday's law and the Maxwell-Ampere law and \eqref{longitudinal_components} and performing more iterations of the nonparaxial approximation of the beam, $W_{1-3}/W_0$ should approach the same values as were calculated in Fig.~4 of the main article.
That said, Fig~\ref{figS2} confirms the importance of longitudinal field components for generating a particle-like EM ellipse texture.

\bibliography{bibliography}